\documentclass{emulateapj}
\usepackage{graphicx}
\usepackage{natbib}

\newcommand{\sn}{$n$}
\newcommand{\mie}{$\langle\mu\rangle_{e}$}

\newcommand{\nablagr}{$\nabla_{g-r}$}

\newcommand{\nablagX}{$\nabla_{g-X}$}
\newcommand{\gX}{$g-X$}
\newcommand{\gXc}{$(g-X)_{0.1}$}
\newcommand{\grc}{$(g-r)_{0.1}$}
\newcommand{\gKc}{$(g-K)_{0.1}$}

\newcommand{\nablat}{$\nabla_t$}
\newcommand{\nablaz}{$\nabla_Z$}
\shorttitle{ }
\shortauthors{ }

\begin{document}
\title{  On  the Radial  Stellar  Content of  Early-Type  Galaxies as  a Function of Mass and Environment}

\author{F.  La  Barbera\altaffilmark{1}, I.  Ferreras\altaffilmark{2},
  R.R. de Carvalho\altaffilmark{3}, P.A.A. Lopes\altaffilmark{4}, A. Pasquali\altaffilmark{5}, I.G. de la Rosa\altaffilmark{6, 7, 8}, G. De Lucia\altaffilmark{9} }
\affil{$^{(1)}$INAF -- Osservatorio Astronomico di Capodimonte, Napoli, Italy }
\affil{$^{(2)}$MSSL, University College London, Holmbury St Mary, Dorking, Surrey RH5 6NT, UK}
\affil{$^{(3)}$Instituto Nacional de Pesquisas Espaciais/MCT, S. J. dos Campos, Brazil}
\affil{$^{(4)}$Observat\'orio do Valongo/UFRJ, Rio de Janeiro, Brazil}
\affil{$^{(5)}$Astronomisches Rechen Institut, Zentrum f\"ur Astronomie der Universit\"at Heidelberg, M\"onchhofstr. 12--14, 69120 Heidelberg, Germany}
\affil{$^{(6)}$Instituto de Astrof\'\i sica de Canarias (IAC), E-38200 La Laguna, Tenerife, Spain}
\affil{$^{(7)}$Department of Physics and Astronomy, University College London, Gower Street, London, WC1E 6BT}
\affil{$^{(8)}$Departamento de Astrof\'\i sica, Universidad de La Laguna, E-38205, Tenerife, Spain}
\affil{$^{(9)}$INAF -- Osservatorio Astronomico di Trieste, via G.B. Tiepolo 11, 1-34143 Triste, Italy}

\begin{abstract}
Using optical--optical and optical--NIR colors, we analyze the radial
dependence of age and metallicity inside massive ($ M_\star \gtrsim
10^{10.5} \, M_\odot$), low-redshift ($z<0.1$), early-type galaxies
(ETGs), residing in both high-density group regions and the field.  On
average, internal color gradients of ETGs are mainly driven by
metallicity, consistent with previous studies.  However, we find that
group galaxies feature positive age gradients, \nablat, i.e. a younger
stellar population in the galaxy center, and steeper metallicity
gradients, compared to the field sample, whose \nablat \, ranges from
negative in lower mass galaxies, to positive gradients at higher mass.
These dependencies yield new constraints to models of galaxy formation
and evolution.  We speculate that age and metallicity gradients of
group ETGs result from (either gas-rich or minor-dry) mergers and/or
cold-gas accretion, { while field ETGs exhibit the characteristic
  flatter gradients expected from younger, more metal-rich, stars
  formed inside--out by later gas-cooling.}
\end{abstract}


\section{Introduction}

Being hot dynamical systems, ETGs  are often assumed to be formed from
major mergers, where  the progenitors can be either  gas-rich disks or
ETGs.   The  predominantly  old  stellar populations  found  in  these
galaxies indicate a rapid and intense star formation history, although
the detection of compact,  massive ETGs at high redshift \citep{dad05}
implies a  change in size  of a factor  3--5 between z$\sim$2  and z=0
\citep[see e.g.][]{truj07}.

Recent studies  of the  size evolution of  massive ETGs over  the past
8~Gyr propose a number of alternative scenarios, such as minor mergers
(\citealt{naab09};  \citealt{truj11}) or a  puffing-up of  the central
regions  caused  by  the  baryonic  mass  loss  \citep{dam09,  fan10}.
Alternatively, cold accretion -- perhaps reminiscent of the monolithic
collapse scenario \citep{lar75} --  may provide an independent channel
for  the formation  of  massive galaxies,  accounting  for the  highly
efficient   star  formation   measured  in   these  systems   at  high
redshift~\citep{dekel09}.  In addition, the environment can contribute
as well  to the star formation  history.  As a galaxy  enters a group,
its  hot gas  reservoir can  be removed,  shutting off  star formation
(``strangulation'').   All these processes  { may
  be discerned}  if spatial information within galaxies  is taken into
account.  The  majority of  studies on the  star formation  history of
ETGs  use  observations  integrated  within an  aperture,  losing  the
discriminating  power.  Radial  gradients of  photo-spectroscopic data
represent  a  fundamental  observable  to  disentangle  the  different
formation channels  described above. Studies of local  samples of ETGs
show   that   the  majority   feature   red   {  cores   \citep[see
    e.g.][]{Peletier:90}}.  Blue cores are less frequent, being mostly
low-mass    ($M_\star   \widetilde{<}   10.^{10.5}    M_\odot$)   ETGs
\citep{suh10}.  Radial  gradients of spectral line  strengths reveal a
significant      trend     in     {      metallicity     \citep[see
    e.g.][]{forbes05,ogando05,spolaor09,koleva11} }  and an intriguing
dependence  of   the  age  gradients   on  environment  \citep{psb06}.
Extending  the analysis to  moderate redshift  takes advantage  of the
lookback  time,  as  younger  populations  are less  affected  by  the
age-metallicity degeneracy.   The analysis  of the color  gradients of
the  GOODS sample  of ETGs  over a  wide redshift  range (0.4$<$z$<$1)
confirmed  a  clear metallicity  radial  gradient  for  the red  cored
galaxies, and a significantly younger population at the centers of the
blue-cored  galaxies  \citep{Ferreras:09},  with  an increase  in  the
fraction of blue- vs. red-cored profiles towards lower stellar masses.

Galaxy mass and environment are  the two main drivers of the formation
history  of galaxies \citep[see  e.g.  ][]{wei09}  and recent  work on
ETGs  has revealed  a  significant difference  between  the two,  with
environment only  playing a secondary role on  the stellar populations
\citep{benv10}  although with  more significant  effects  on dynamical
relationships  such  as the  {  Fundamental Plane  \citep[hereafter
    LLD10]{PaperIII}  }.  In this  letter, we  extend the  analysis by
inferring age  and metallicity  radial gradients of  low-redshift ETGs
from a multi-wavelength  dataset in the optical and  NIR. Our findings
reveal a significant contribution  of the environment to the formation
and assembly mechanisms of ETGs.

\section{SAMPLE}
\label{sec:samples}

The SPIDER survey consists of a volume-limited sample of $39,993$ {\it
  bright}  ETGs ($M_r<-20$)  in the  redshift  range $0.05  \le z  \le
0.095$,  with  $griz$  photometry  from SDSS-DR6.   $5,080$  of  these
galaxies  also have  $YJHK$ imaging  from UKIDSS-Large  Area Survey~({
  see~LLD10, and references therein,  for details on data analysis and
  sample selection}).  { ETGs are defined as galaxies with a prominent
  bulge\footnote{accounting  for  $>80\%$  of  total  light,},  and  a
  passive spectrum in their  central regions (within the SDSS fiber)}.
Galaxies  have structural parameters,  i.e.  effective  radius, $R_e$,
mean surface  brightness within that  radius, \mie, and  Sersic index,
\sn, homogeneously measured  from $g$ through $K$ with  the software {
  2DPHOT, by}  fitting galaxy  images with seeing-convolved  2D Sersic
models.  The environment of ETGs  is characterized by an FoF catalogue
of $8,083$ groups.  { A  shifting gapper technique is applied to} this
catalogue (see~\citealt{lop09a}) allowing galaxies to be classified as
either  group  members ($\sim46\%$),  {  non-group members  (hereafter
  ``field'' galaxies}; $\sim 33 \%$), or unclassified ($\sim 21 \%$; {
  see~LLD10}).  Each  group galaxy  has 2D local  density, $\Sigma_N$,
estimated by its distance to  the n-th nearest group member, where $n$
scales as the square root of group richness.

We select only galaxies with better quality structural parameters from
$g$   through    $K$   ($\sim   90    \%$   of   the    sample;   {
  see~\citealt[hereafter LDD10]{PaperIV}}),  and stellar mass $M_\star
\ge  3 \times  10^{10} M_{\odot}$,  which gives  an $M_\star$-complete
sample of  ETGs\footnote{The $M_\star$ limit is  established with the
  same approach as in fig.~3 of { LDD10}.}. These selections result
into a sample of 9,285 (936) field galaxies with optical (optical+NIR)
data available.  For  group galaxies, we select only  ETGs residing in
the highest  density regions,  $\Sigma_N \gtrsim 16.6  {\rm gals}/{\rm
  Mpc}^2$  ($\Sigma_N \gtrsim  6.4  {\rm gals}/{\rm  Mpc}^2$) for  the
optical (optical+NIR) sample.   These $\Sigma_N$ thresholds are chosen
to emphasize  the environmental dependence of  galaxy color gradients,
without overly  reducing the number  of group galaxies\footnote{Since
  optical--NIR  color   gradients  do  not   depend  significantly  on
  $\Sigma_N$ (Sec.~\ref{sec:cgs_mass_env}),  using the same $\Sigma_N$
  threshold for both optical  and optical+NIR samples would not change
  our results.  However, adopting a lower $\Sigma_N$ threshold for the
  optical+NIR  sample allows  us to  effectively reduce  error bars.},
which  consist of $1,792$  (optical) and  $487$ (optical+NIR)  ETGs. {
  Most of  the parent systems  where these ETGs reside  are relatively
  poor groups, with an average  velocity dispersion of $\sim 370 \, km
  \,  s^{-1}$}.  Both  field  and  group ETGs  are  split among  three
$M_\star$-bins:  ``low'' ($3  {<}  M_\star  \!  {<}  5  \!  \times  \!
10^{10}  M_\odot$), ``intermediate''  ($5  {<} M_\star  \!   {<} 8  \!
\times \!  10^{10} M_\odot$), and  ``high'' ($8 {<} M_\star \!  {<} 45
\!  \times \!   10^{10} M_\odot$), being the number  of group ETGs the
same in each bin.


\section{Internal colors vs. galaxy mass and environment}
\label{sec:cgs_mass_env}

The internal  color gradient of  an ETG, defined  as the slope  of its
radial color profile, written as $\nabla_{g-X}$, where $X=rizYJHK$, is
estimated as described in { LDD10}, using deconvolved Sersic models
of each  galaxy to  measure $g-X$ colors  on concentric  ellipses with
axis ratio and position angle fixed  to the r-band Sersic fit.  A line
is fitted to the resulting $g-X$  vs.  $\log R$ profile, in the radial
range of $0.1$ to  $1 R_e$ (as in, e.g.,~\citealt{Peletier:90}), where
$R$ is the equivalent galactocentric  distance and $R_e$ is the r-band
effective radius. The line slope gives \nablagX.

\begin{figure}[t!]
\begin{center}
\includegraphics[height=125mm]{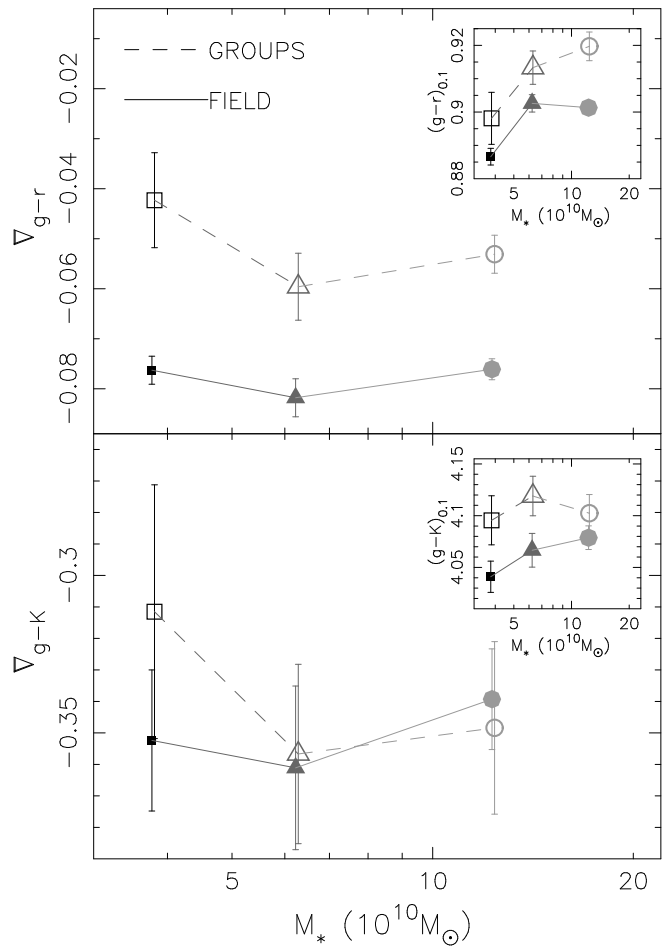}
\caption{ Color  gradients of  ETGs as a  function of  environment and
  stellar  mass.  Group  and field  ETGs  are plotted  with empty  and
  filled  symbols, respectively.   Low-, intermediate-,  and high-mass
  galaxies are plotted with squares, triangles, and
  circles,  respectively.  Upper  panel: median  $\nabla_{g-r}$  versus
   median  stellar   mass.   Error  bars  are  1~$\sigma$
  confidence intervals  on median values. The inset  shows how central
  values  of  $g-r$,  $(g-r)_{0.1}$,   depends  on  stellar  mass  and
  environment.   Lower panel:  the same  as  upper panel  but for  the
  optical--NIR color $g-K$.
\label{fig:col_mass_env}
}
\end{center}
\end{figure} 

Fig.~\ref{fig:col_mass_env}    shows   the   dependence    of   median
optical--optical and optical--NIR  color gradients, $\nabla_{g-r}$ and
$\nabla_{g-K}$\footnote{The  other  available  colors  ($g-i$,  $g-z$,
  $g-Y$, $g-J$, and $g-H$) are not shown for brevity, but are included
  in the stellar  population analysis of Sec.~\ref{sec:tz_mass_env}.},
as well  as central  colors, \grc  \, and \gKc  { (computed  at $R=0.1
  R_e$)}, on galaxy mass and environment (i.e.  field vs.  group).  To
a first  approximation, optical--optical colors are  more sensitive to
the  age   of  a  stellar  population  than   its  metallicity,  while
optical--NIR   colors   are  more   sensitive   to  metallicity   than
age~\citep[see e.g.][]{deJong:96}.  Hence,  the upper and lower panels
of Fig.~\ref{fig:col_mass_env}  can be roughly seen  as reflecting the
behavior  of age and  metallicity, respectively.   At a  given stellar
mass,  group galaxies  have  significantly shallower  optical--optical
color  gradients than  their field  counterparts\footnote{  We note
  that this  trend cannot  be caused by  an increase  of the E  vs. S0
  fraction  in the  group environment,  as lenticulars  have shallower
  color     gradients    (e.g.~\citealt{Roediger:11}).},    consistent
with~\citet{LdC:05}, who found cluster galaxies to have systematically
flatter  \nablagr \,  (by $\sim  0.03 \,  {\rm mag}/{\rm  dex}$  at $z
\lesssim 0.1$)  with respect to  galaxies in less  dense environments.
However, optical--NIR  color gradients do not  exhibit any significant
environmental dependence, implying  that age (rather than metallicity)
gradients  are changing  with the  environment.  The  small  panels in
Fig.~\ref{fig:col_mass_env}  also  show  that,  at all  masses,  group
galaxies  have redder  central colors  than  those in  the field.   On
average, the color difference amounts to $\sim 0.02$~mag in $g-r$, and
$\sim  0.05$~mag  in  $g-K$.   This  may  result  from  both  age  and
metallicity increasing  with mass, and group (relative  to field) ETGs
having older stellar populations, as found by, e.g.,~\citet{Thomas:05,
  Gallazzi:06, Pasquali:10}.


\section{Age and metallicity gradients vs. galaxy mass and environment}
\label{sec:tz_mass_env}
We assume that  the dependence of internal colors of  ETGs on mass and
environment  is  mainly driven  by  the  age  and metallicity  of  the
underlying stellar populations, which is motivated by the existence of
well-established    absorption-line    gradients    in   ETGs    (see,
e.g.,~\citealt{SFS:07, RSL:10}).   Although a dust  component may also
generate color gradients~\citep{SW:96},  no observational evidence has
been  found   so  far     that  dust   can  play  a   major  role
(\citealt{SWF:09}; { LDD10}).

\begin{figure}[t!]
\begin{center}
\includegraphics[height=125mm]{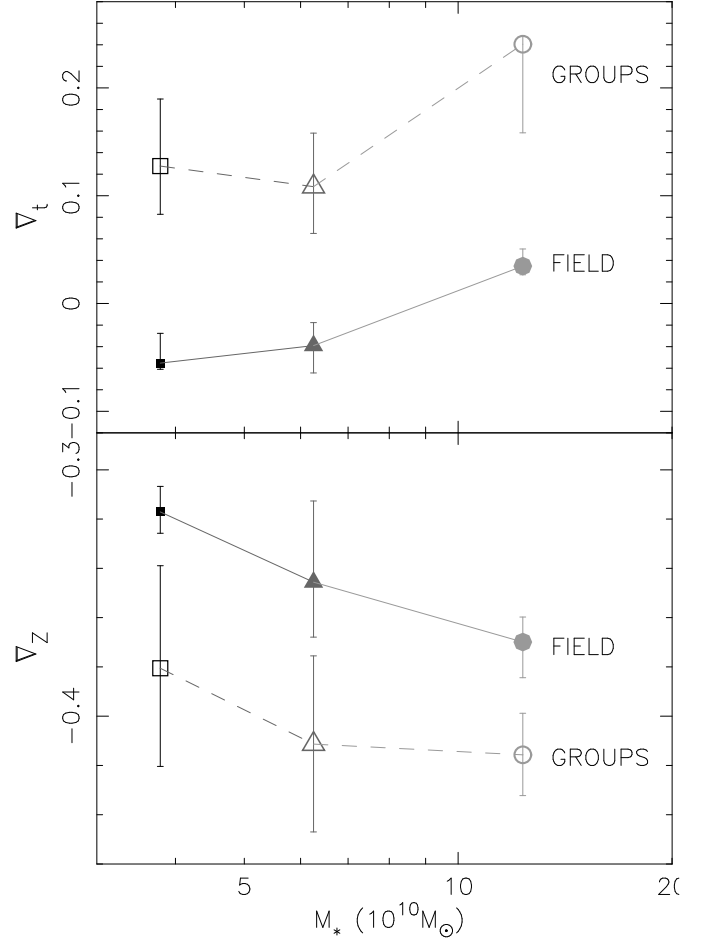}
\caption{ Age (upper-panel) and metallicity (lower-panel) gradients of
  ETGs as a function of stellar mass and environment.  Symbols are the
  same  as   in  Fig.~\ref{fig:col_mass_env}.  Error   bars,  defining
  1~$\sigma$ confidence intervals, are obtained by shifting the \gXc's
  and  \nablagX's according to  their errors,  and repeating  the fit.
\label{fig:tz_mass_env}
}
\end{center}
\end{figure} 

We fit simultaneously all available central colors and color gradients
(as shown in Fig.~\ref{fig:col_mass_env}) with stellar population (SP)
models, inferring galaxy central age and metallicity\footnote{ Central
  age and metallicity  are not discussed here, as  their estimates are
  less robust than that of SP gradients, involving an absolute, rather
  than a relative, matching of  model to observed colors.}, as well as
age  and   metallicity  gradients~\footnote{i.e.   The   variation  of
  logarithmic  age  and   metallicity  per  decade  in  galactocentric
  distance.},    \nablat     \,    and    \nablaz     \,    (as    in,
e.g.,~\citealt{Tortora:10}).   Each SP  model  consists of  a pair  of
Simple Stellar  Populations (SSPs) with~\citet{Chabrier:03}  IMF, from
the Charlot  \& Bruzual (2011,  in preparation; CB11)  synthesis code.
The ``inner''  SSP is used to  model central galaxy  colors, while the
difference  of \gX  \, between  the two  SSPs is  fitted  to \nablagX,
providing \nablat  \, and \nablaz. The  fitting is done  in a $\chi^2$
sense  by  varying age  and  metallicity of  both  SSPs  in the  range
$1<t<14$~Gyr  and  $0.2<Z/Z_\odot<2$,  respectively.   We  found  that
similar  results  are  obtained  when  using~\citet{BrC03}  population
synthesis  models,  or  using exponentially  declining  star-formation
models.    As  shown  in   Fig.~\ref{fig:tz_mass_env},  the   age  and
metallicity content of ETGs  depends significantly on both galaxy mass
and  environment,  reflecting  the  behavior of  optical--optical  and
optical--NIR   colors   of   Fig.~\ref{fig:col_mass_env}.    For   all
environments and mass bins, the main driver of color gradients in ETGs
is   a  {  significantly   negative  ($<-0.3$)   metallicity  gradient
  (consistent  with  e.g.~\citealt{PVJ90,  Ferreras:09}),  with  field
  galaxies having somewhat shallower  \nablaz \, (by $\sim 0.06$) than
  their  group  counterparts}.   However,  age  is  found  to  play  a
significant role,  and, perhaps more importantly, this  depends on the
environment  where  galaxies reside.   Group  galaxies have  positive,
albeit small  , age gradients ($\sim  0.1$, i.e.  an  age variation of
$\sim 23\%$  per radial decade),  with the galaxy  stellar populations
being younger  in the center than  in the outskirts,  while field ETGs
exhibit  systematically  smaller,   mostly  negative,  age  gradients,
consistent  with  \citet{Ferreras:09} based  on  the  analysis of  the
evolution  of the  color  gradients  of field  ETGs  from GOODS  data.
Notice, however,  that at high-mass  field galaxies also tend  to have
slightly positive \nablat ($\sim 0.04 \pm 0.01$).




\section{Discussion}
\label{sec:discussion}





{ The analysis presented in this Letter suggests that there is a
  significant trend of the internal distribution of age and
  metallicity in massive ETGs with galaxy mass and environment. The
  results of Fig.~\ref{fig:tz_mass_env} can be qualitatively discussed
  in the light of current formation scenarios of ETGs, considering
  that most ($\sim 90\%$) of our sample of group galaxies consists of
  satellite, rather than central, galaxies (according to the
  definition of~\citealt{Yang:07}):

{\it - Gas-rich mergers.}  The presence of a younger stellar
population (i.e. a positive age gradient) in the center of group ETGs
supports a dissipative formation picture, whereby gas-rich mergers
fuel the central region with cold gas, allowing for the formation of
younger, metal-enriched, stars.  In a hierarchical picture of galaxy
formation, the more massive ETGs start forming stars at earlier epochs
(albeit assembling later, see~\citealt{deLucia:06}), when disk-like
progenitors likely had a more turbulent and clumpy
ISM~\citep{SINS:09}, possibly implying a larger amount of dissipation.
This might explain why age (metallicity) gradients tend to marginally
increase (decrease) with stellar mass.  However, this is not supported
by SPH simulations of galaxy formation, with merging producing no
correlation of metallicity gradients and mass~\citep{Kobayashi:04}.
Moreover, both the light profile shape as well as the size--mass
relation of ETGs at $z \sim 0$ provide evidence against an increasing
importance of gas dissipation at high
mass~\citep{Hopkins:09,Shankar:11}.

{\it - Cold accretion.} An alternative scenario to explain positive
age gradients involves the accretion of cold (clumpy) gas from the
cosmological surroundings, leading to subsequent star formation in the
central region. Cold accretion is expected to be more important at
high redshift ($z>2$), being a relevant mode for the formation of
galaxies as massive as $\sim 10^{12} M_\odot$~(\citealt{dekel09}, but
see also~\citealt{Keres:05}).  In a cold accretion scenario, where the
formation process resembles monolithic collapse, the \nablat \, and
\nablaz \, would naturally correlate with galaxy mass as star
formation lasts longer in the center of more massive systems having a
deeper central potential well (e.g.~\citealt{Kobayashi:04,
  Pipino:10}).  This interpretation is also supported by the fact
that: (1) the ages and metallicities of the stellar populations within
and among ETGs are tightly correlated to the local escape
velocity~\citep{Scott:09}; and (2) the resulting gradients depend on
mass (e.g.~\citealt{forbes05}) and, at fixed mass, on the duration of
star-formation, a property that can be parametrized by the
$\alpha$ to iron abundance ratio ({ LDD10}).  One should notice,
however, that the existence of a correlation between \nablaz \, and
mass has been longly debated in the literature (see, e.g.,
~\citealt{spolaor09, koleva11}, and references therein).  

{\it  - Later  gas accretion.}  While group  galaxies have  their star
formation    quenched   when    entering    bigger   halos    (through
``strangulation''),  field   galaxies  can  accrete   gas  longer,  by
radiative cooling  of their dark-matter hot-gas reservoir,  and form a
younger,  more  metal-rich, stellar  component  outwards.  This  would
lower,  and   eventually  invert,   the  age  gradient,   making  also
metallicity gradients shallower, consistent with what we see for field
ETGs in Fig.~\ref{fig:tz_mass_env}.

{\it - Stripping.} Environmental  effects, such as ram pressure and/or
tidal stripping, may also trigger (central) star formation in recently
accreted  group  satellites  (e.g.~\citealt{Bekki:03}).   In  general,
stripping   should   be  less   important   for  high-mass   galaxies,
inconsistent  with  the  trends  seen  in  Fig.~\ref{fig:tz_mass_env}.

{\it  - Dry  merging.} ETGs  can also  form by  gas-poor interactions.
Minor dry mergers and stellar accretion can drive the formation of the
outer  envelopes  of  ETGs,   explaining  their  size  evolution  with
redshift~\citep{naab09}.   Gas-poor  mergers  should mix  the  stellar
populations within  galaxies, flattening pre-existing  metallicity and
age  gradients~\citep{White:80}.   Since  \nablat  \, and  \nablaz  \,
steepen  with mass (see  also {  LDD10}), our  data seem  to reject
major  dry-mergers as  the main  channel  for the  formation of  ETGs,
unless the amount of { flattening is small~\citep{Hopkins:09}}.  Minor
dry-mergers  would  lead to  a  deposit  of  metal-poor, old,  stellar
material in the outer regions of a galaxy, increasing (decreasing) the
age (metallicity) gradients.  Therefore,  the trends of \nablat \, and
\nablaz \, in Fig.~\ref{fig:tz_mass_env}  might also result from minor
dry-mergers to  be more important at high-mass  (consistent with model
predictions; see~\citealt{Hopkins:10,  deLucia:11}) and preferentially
in  group, relative  to field,  environments.  However,  such  a trend
would  be  at  odds  with  the  finding of  a  lack  of  environmental
dependence  of  the  redshift  evolution  of  ETGs  on  the  mass-size
plane~\citep{Rettura:10}.

In summary, although we find that several mechanisms can contribute to
the observed age  and metallicity gradients in ETGs  (among them, cold
accretion  and minor  dry mergers  are the  most likely  ones),  we do
indeed detect  physical differences  in the formation  mechanisms with
respect to environment, with field (relative to group) galaxies having
the  characteristic flatter gradients  expected from  inside--out star
formation induced  by late  gas-cooling.  Studies of  radial gradients
outside  of  the  effective  radius  will shed  more  light  into  the
mechanisms that drive the formation of early-type galaxies.

}

\begin{acknowledgements}
We          used          data          from         the          SDSS
(http://www.sdss.org/collaboration/credits.html).  This  work is based
on   data  obtained   as  part   of  the   UKIRT  Infrared   Deep  Sky
Survey~\citep{Law07}.   IGR  acknowledges  a  grant from  the  Spanish
Secretaria General  de Universidades of the Ministry  of Education, in
the  frame  of  its  programme  to promote  the  mobility  of  Spanish
researchers to foreign centers.
\end{acknowledgements}

\end{document}